\RequirePackage{fix-cm}
\documentclass[11pt]{llncs}

\usepackage{fullpage}

\def \proc{{\bf proc}\ }
\def \lact{\mathop{|\joinrel\![}}
\def \ract{\mathop{]\joinrel\!|}}

\def \hvar{h}
\def \hrvar{{\sf hr}}
\def \ttvar{tt}
\def \outtvar{{\sf out_t}}
\def \outIvar{{\sf out_1}}

\def \pctvar{pc_t}
\def \hpctvar{{\widehat{pc}_t}}

\def \calltvar{{\sf call_t}}
\def \callIvar{{\sf call_1}}

\def \endbox{\hfill \qed}

\usepackage{xspace}

\def \follows {\Leftarrow} 

\def \nasgn {\mathbin{:\joinrel\in}}
  
\def \bs{\backslash}

\usepackage{graphicx}
\usepackage{rcp}
\usepackage{url}
\usepackage{amsmath% ,amssymb
}
\usepackage{color}
\usepackage{stmaryrd}
\usepackage{oz}
\usepackage{zed-csp}
\usepackage{wasysym}

\newcommand{\beh}[1]{\llbracket #1 \rrbracket}

\def \diverge {{\mathop \uparrow}}

\def \all  {\forall}
\def \asgn  {{\;:=\;}}

\def \refs  {{\;\sqsubseteq\;}}

\def \sref  {\sqsubseteq}
\def \srefc  {\mathbin{\widehat{\sqsubseteq}}}

\def \var {{\bf var}\ }
\def \rav {{\bf rav}\ }
\def \lvar {{\bf var_u}\ }
\def \gvar {{\bf var_o}\ }

\def \val {{\bf val}\ }
\def \res {{\bf res}\ }

\newcommand{\sskip}{\mathbf{skip}}

\newcommand{\ddo}{\mathbf{do}}
\newcommand{\ood}{\mathbf{od}}

\renewcommand{\Init}{{\bf initialisation}\ }

\usepackage{mathrsfs}
\newcommand{\msA}{\mathscr{A}}
\newcommand{\msB}{\mathscr{B}}
\newcommand{\msC}{\mathscr{C}}

\newcommand{\mcC}{\mathcal{C}}
\newcommand{\mcD}{\mathcal{D}}
\newcommand{\mcP}{\mathcal{P}}
\newcommand{\mcR}{\mathcal{R}}

\newcommand{\mcM}{\mathcal{M}}

\newcommand{\MGC}{\mathcal{M}}

\newcommand{\refeq}[1]{(\ref{#1})}
\newcommand{\refeqn}[1]{(\ref{#1})}

\newcommand{\reffig}[1]{Fig.~\ref{#1}}
\newcommand{\refthm}[1]{Theorem~\ref{#1}}
\newcommand{\reflem}[1]{Lem\-ma~\ref{#1}}

\newcommand{\refsec}[1]{Section~\ref{#1}}
\newcommand{\refex}[1]{Example~\ref{#1}}
\newcommand{\refdef}[1]{Definition~\ref{#1}}

\authorrunning{Brijesh Dongol and Lindsay Groves} 
\titlerunning{Contextual trace refinement for concurrent objects}

\usepackage[scaled]{beramono}
\usepackage[T1]{fontenc}

\newcommand{\OMIT}[1]{{}}%{\color{red}{#1}}}
\usepackage{color}

\newcommand{\invoke}[3]{inv(#1, #2, #3)}
\newcommand{\return}[3]{ret(#1, #2, #3)}
\newcommand{\invokeb}[2]{#2_{#1}}
\newcommand{\returnb}[2]{\widehat{#2}_{#1}}

\title{Contextual trace refinement for concurrent objects: \\Safety
  and progress}

\author{Brijesh Dongol\inst{1} \and Lindsay Groves\inst{2}}

\institute{Department of Computer Science,\\
  Brunel University London, UK\\
 \email{Brijesh.Dongol@brunel.ac.uk} \\[1mm] \and 
 School of Engineering
  and Computer Science, Victoria\\
  University of Wellington, New Zealand\\
  \email{lindsay@ecs.vuw.ac.nz}}

\begin{document}

\maketitle

\begin{abstract}
  Correctness of concurrent objects is defined in terms of safety
  properties such as linearizability, sequential consistency, and
  quiescent consistency, and progress properties such as wait-, lock-,
  and obstruction-freedom. These properties, however, only refer to
  the behaviours of the object in isolation, which does not tell us
  what guarantees these correctness conditions on concurrent objects
  provide to their client programs.  This paper investigates the links
  between safety and progress properties of concurrent objects and a
  form of trace refinement for client programs, called contextual
  trace refinement.  In particular, we show that linearizability
  together with a minimal notion of progress are sufficient properties
  of concurrent objects to ensure contextual trace refinement, but
  sequential consistency and quiescent consistency are both too
  weak. Our reasoning is carried out in the action systems framework
  with procedure calls, which we extend to cope with non-atomic
  operations.
\end{abstract}

  % This enables one to reason
  % about trace refinement of a client program that uses a concurrent
  % object in place of a corresponding abstract specification. 
  % However, sequential consistency, and
  % quiescent consistency are too 

\section{Introduction}

Concurrent objects provide operations that can be executed
simultaneously by multiple threads, and provide a layer of abstraction
to programmers by managing thread synchronisation on behalf of client
programs, which in turn improves safety and efficiency. Correctness of
concurrent objects is usually defined in terms of
%a sequential specification
%object, however, because several operations may be executed
%simultaneously, correctness of operations cannot be stated in terms of
%pre/post-conditions. Instead, they are defined in terms of 
the possible \emph{histories} of invocation and response events generated by
executing the operations of a sequential specification object.  
There are several notions of
safety for concurrent objects \cite{HeSh08,DDGS15-ECOOP}: sequential
consistency, linearizability, and quiescent consistency being the most
widely
used.
%Concurrent objects provide operations that can be executed
%simultaneously by multiple threads, and provide a layer of abstraction
%to programmers, managing thread synchronisation on behalf of their
%clients to improve safety and efficiency. Correctness of concurrent
%objects is usually defined in terms of a sequential specification
%object, however, because several operations may be executed
%simultaneously, correctness of operations cannot be stated in terms of
%pre/post-conditions. Instead, they are defined in terms of the
%possible \emph{histories} of invocation/response events generated by
%executing the operations of the object.  There are several notions of
%safety for concurrent objects \cite{HeSh08,DDGS15-ECOOP}: sequential
%consistency, linearizability, and quiescent consistency being the most
%widely
%used.
% These define a mapping from a history $h$ of a concurrent
% implementation to a sequential history --- $h$ is deemed to be valid
% if some sequential history resulting from the mapping is a possible
% history of the sequential specification object.
Similarly, there are many different notions of progress
\cite{HeSh08,HS11}, e.g., wait-, lock- and obstruction-freedom are popular non-blocking
conditions. % , that may be used to describe the trace of a
% concurrent object.
% There are many different techniques for expressing
% these --- we use a linear temporal logic encoding over the histories
% of an object \cite{MP92}.

Both safety and progress properties are stated in terms of a
concurrent object in isolation, and disregard their context, i.e., the
client programs that use them.  Programmers (i.e., client developers)
have therefore relied on informal ``folk theorems'' to link
correctness conditions on concurrent objects and substitutability of
objects within client programs.  We seek to provide a formal account
of this relationship, addressing the question: ``Provided concurrent
object $OC$ is correct with respect to sequential object $OA$, how are
the behaviours of $\mcC[OA]$ related to those of $\mcC[OC]$?'', where
$\mcC[O]$ denotes a client program $\mcC$ that uses object $O$, for
different notions of correctness.  One of the first formal answers to
this question was given by the \emph{abstraction theorems} of
Filipovi\'c et al.\ \cite{FORY10}, who link safety properties
sequential consistency and linearizability to a contextual notion of
correctness called \emph{observational refinement}, which defines
substitutability with respect to the initial and final state of a
system's
execution. % \footnote{Observational refinement is equivalent to the
  % notion of data refinement by He et al.\ \cite{HeHS86}.}
For \emph{terminating} clients, linearizability is
shown to be equivalent to observational refinement, while sequential
consistency is shown to be equivalent to observational refinement provided
that clients only communicate via shared concurrent objects.

Since non-termination is common in many concurrent systems, e.g.,
operating systems and real-time controllers, our work aims to
understand substitutability for \emph{potentially non-terminating}
clients.  Related to this aim is the work of Gotsman and Yang
\cite{GY11} and Liang et al.\ \cite{LiangHFS13}, who link
observational refinement to safety and progress properties of
concurrent objects.
% they also use a more liberal definition of linearizability, where
% abstract specifications are potentially non-atomic.
However, both \cite{GY11} and \cite{LiangHFS13} assume that the
concurrent objects in question are already linearizable. Furthermore,
\cite{GY11} aims to understand \emph{compositionality} of progress
properties, while \cite{LiangHFS13} develops \emph{characterisations}
of progress properties based on the observational guarantees they
provide.
 % The overall effect is that this recasts
% % Filipovi\'c et al.'s somewhat
% the results of \cite{FORY10,GY11,LiangHFS13} in an action systems
% setting.  

The motivation for our work differs from \cite{GY11,LiangHFS13} in
that we take \emph{contextual trace refinement} as the underlying
correctness condition when substituting $OC$ for $OA$ in $\mcC$, then
aim to understand the safety/progress properties on $OC$ that are
required to guarantee trace refinement between $\mcC[OA]$ and
$\mcC[OC]$. To this end, we develop an \emph{action systems} framework
that integrates and extends existing work \cite{SereW00,back94trace}
from the literature, building on our preliminary results on this topic
\cite{DG15-REFINE}.  As part of our contributions we (i) extend Sere
and Wald\'en's treatment of action systems with procedures
\cite{SereW00} with \emph{non-atomic procedures}; (ii) develop a
theory for \emph{contextual trace refinement}, adapting Back and von
Wright's \cite{back94trace} theory for trace refinement of action
systems, then reduce system-wide proof obligations (i.e., properties
of the client and object together) to proof obligations on the objects
only; % (iv) motivate the safety and progress properties required of
% concurrent objects to ensure contextual trace refinement; 
(iii) show that linearizability \cite{Herlihy90} and minimal progress
\cite{HS11} together are sufficient to guarantee contextual trace
refinement; and (iv) show that both sequential consistency and
quiescent consistency are too weak for contextual trace refinement,
even when client threads only communicate through the shared
object.

\vspace{-2mm}

% \begin{figure}[t]
%   \centering
%   \scalebox{0.6}{\input{proof-chain.pspdftex}}
%   % \vspace{-2mm}
%   \caption{Dashed arrows denote theory instantiation and solid arrows
%     denote implication}
%   \label{fig:proof-chain}
%   \vspace{-2mm}
% \end{figure}

\section{Concurrent objects and their clients}
\label{sec:concurrent-objects}

% We motivate our work using the widely-used non-blocking stack by
% Treiber (\refsec{sec:exampl-treib-stack}), which we use informally to
% discuss correctness. An example stack client
% (\refsec{sec:concurrent-clients}) is used to informally motivate
% contextual trace refinement. % Correctness conditions and trace
% % refinement are formalised in Sections \ref{sec:refin-client-object}
% % and \ref{sec:trace-refin-client}, respectively.

\vspace{-1mm}

\subsection{Client-object systems}
\label{sec:exampl-treib-stack}

We consider concurrent systems where a client consists of multiple
threads which interact with one or more concurrent objects and shared
variables.  For example, the following client program consists of threads
{\tt 1} and {\tt 2} using a shared stack {\tt s}, and variables {\tt x}, 
{\tt y} and {\tt z}.\smallskip

  \hfill  \begin{minipage}[b]{0.9\linewidth}
    \small\tt Init x, y, z = 0, 0, 0
      \\
      \begin{minipage}[t]{0.4\columnwidth}
        \tt Thread 1:
        
        \ \ T1:\ s.push(1);
        
        \ \ T2:\ s.push(2);
        
        \ \ T3:\ s.pop(x);
        
        % \ \ T4:\ x := \outt ;
        
      \end{minipage}
      \hfill
      \begin{minipage}[t]{0.4\columnwidth}
        \tt Thread 2:
        
        \ \ U1:\ s.pop(y);
        
        % \ \ U2:\ y := \outu ;
        
        \ \ U2:\ z := x;
      \end{minipage}
    \end{minipage}      \smallskip

  \noindent
  Thread {\tt 1} pushes {\tt 1} then {\tt 2}
  onto the stack {\tt s}, then pops the top element of {\tt s} and
  stores it in {\tt x}. Concurrently, thread {\tt 2} pops the top
  element of {\tt s} and stores it in {\tt y}, then reads the value of
  {\tt x} and stores it in {\tt z}.

The abstract behaviour of a stack is defined in terms of a sequential
object, as shown in \reffig{fig:Abstract-TS}. The abstract stack consists
of a sequence of elements $S$ together with two operations $push$ and $pop$
(`$\langle$' and `$\rangle$' delimit sequences, `$\emptyseq$' denotes the
empty sequence, and `$\cat$' denotes sequence concatenation). Note that
when the stack is empty, $pop$ returns a special value {\tt empty} that
cannot be pushed onto the stack.

\begin{figure}[t]
  % \rule{\textwidth}{1pt}
  \begin{minipage}[b]{0.31\columnwidth}
    {\tt \small Init: S = $\emptyseq$}\smallskip

    {\tt \small push(v) == 
      
      \ \textbf{atomic} \{ S := $\langle$v$\rangle \cat$S \}}\smallskip

    {\tt \small pop ==
      
      \textbf{atomic} \{ 
      
      \ \ \textbf{if} S = $\emptyseq$ 
      
      \ \ \textbf{then} \textbf{return} empty
      
      \ \ \textbf{else} 
      
      \ \ \ \ lv := head(S); 

      \ \ \ \ S := tail(S);

      \ \ \ \ \textbf{return} lv 
       \}
    }
    \vspace{-2mm}
    \caption{Abstract stack}
    \label{fig:Abstract-TS}
  \end{minipage}
  \hfill 
  \begin{minipage}[b]{0.62\columnwidth}
    \smallskip
    
    {\tt \small Init: Head = null}\smallskip

    \begin{minipage}[t]{0.46\columnwidth}
      \tt \small push(v) ==
      
      H1:\      n := \textbf{new}(Node);
      
       H2:\ 
      n.val := v;
      
      \ \ \ \ \textbf{repeat}
      
       H3:\ 
      \ \ ss := Head;
      
       H4:\ 
      \ \ n.next := ss;
      
       H5:\ 
      \textbf{until} 

      \ \ \ \ \ \ CAS(Head,ss,n)
      
       H6:\ 
      \textbf{return}
    \end{minipage}
    \hfill
    \begin{minipage}[t]{0.49\columnwidth}
      \tt \small pop ==
      
       \ \ \ \
      \textbf{repeat }
      
       P1:\ ss := Head;
      
       P2:\ \textbf{if} ss = null 
      
      P3:\ \textbf{then} 
 \textbf{return} empty
      
      \ \ \ \ \textbf{else} 
      % \ \ \ \ \ \ \ \textbf{endif};
      
       P4:\ 
      \ \ ssn := ss.next;
      
       P5:\ 
      \ \ lv := ss.val
      
       P6:\ 
      \textbf{until} 
      
      \ \ \ \ CAS(Head,ss,ssn);
      
       P7:\ 
      \textbf{return} lv
    \end{minipage}
    \vspace{-2mm}
    \caption{The Treiber stack}
    \label{fig:TS}
  \end{minipage}
% \rule{\textwidth}{1pt}
  \vspace{-6mm}
\end{figure}

If concurrent objects are implemented using fine-grained concurrency, the
call statements in their clients are not necessarily atomic because they
may invoke non-atomic operations.  Furthermore, depending on the
implementation of {\tt s}, we will get different traces of the client
program because the effects of the concurrent operations on {\tt s} may
take effect in different orders.
For example, \reffig{fig:TS} presents a
simplified version of a non-blocking stack 
example due to Treiber \cite{Tre86}.
%, which has become a standard case
%study from the literature.% \footnote{We assume garbage collection to
  % avoid the so-called ABA problem, where changes to shared pointers
  % may go undetected due to the value changing from some value $A$ to
  % another value $B$ then back to $A$.}
In this implementation, each line of the {\tt push} and {\tt pop}
corresponds to a single atomic step, except {\tt H1}, which may
be regarded as being atomic because a thread can signal to other
threads that a node has been taken in a single atomic
step. % allocating
% a new node is local to the executing thread. 
Synchronisation of {\tt push} and {\tt pop} operations is achieved
using a compare-and-swap (\texttt{CAS}) instruction, which takes as
input a \emph{(shared) variable} {\tt gv}, an \emph{expected value}
{\tt lv} and a \emph{new value} {\tt nv}:\smallskip

\begin{minipage}[t]{0.9\columnwidth}
  \small \tt CAS(gv, lv, nv) $\sdef$ \textbf{atomic} \{
  \begin{tabular}[t]{@{}l@{}}
    \tt \textbf{if} (gv = lv) 
    \tt \textbf{then} gv := nv ; \textbf{return} true \\
    \tt \textbf{else} \textbf{return} false \}
  \end{tabular}
\end{minipage}\smallskip

With this stack implementation, the executions of, say {\tt T1} and {\tt
  U1}, in the above client may overlap, and different behaviours may
be observed according to the order in which steps of the different
threads are executed.  Treiber's stack is linearizable with respect to
the abstract stack in \reffig{fig:Abstract-TS}, so the effect of each
operation call takes place between its invocation and its response.
If a different stack implementation is used which satisfies a more
permissive correctness condition, such as sequential consistency or
quiescent consistency \cite{HeSh08}, a wider range of behaviours may
be observed.

\vspace{-1mm}

\subsection{Observability and contextual trace refinement}

% \fbox{Explain more clearly the difference between observations of a
%   client --- in particular end to end refinement vs trace refinement.}

% How does one judge correctness of a system
% consisting of both a client and the objects it uses? In particular,

With an example client-object system in place, we return to the main
question for this paper: What guarantees do correctness conditions on
concurrent objects provide to clients that use the objects?
Furthermore, how can one address divergence, termination and
reactivity of a client? To address these, we first pin down the
aspects of the system being developed that are visible to an external
observer. Following Filipovi\'{c} et al.\ \cite{FORY10}, we take the
state of the client variables to be observable, and the state of the
objects they use to be
unobservable. % the observable state to be the state of the client
  % variables, and the unobservable state to be the state of the objects
  % they use
Therefore, for the client program in \refsec{sec:exampl-treib-stack},
variables {\tt x}, {\tt y} and {\tt z} are observable, but none of the
variables of the stack implementation {\tt s} are observable. This
allows us to reason about a client with respect to different
implementations of {\tt s}.  Second, we define \emph{when} a system
may be observed. Unlike Filipovi\'{c} et al.\ \cite{FORY10} who only
observe the state at the beginning and end of a client's execution, we
assume that the states \emph{throughout} a client's execution are
visible. This allows us to accommodate, for example, reactive clients,
which interact with an observer in some way even if they are
potentially non-terminating.

Therefore, our notion of correctness for the combined system will be a
form of \emph{observational refinement} that holds iff every
(observable) trace of a client using a concurrent object is equivalent
to some (observable) trace of the same client using the corresponding
abstract specification of the object. The end result is that from the
perspective of a client program, it will be impossible to tell whether
it is using the concurrent object, or its abstract (sequential)
specification.
% We would therefore like to understand how a correctness
% condition between concurrent objects affects trace refinement for
% clients that use these objects.
\begin{example}
  \label{ex:2}
  Let $\mcD$ denote the client program in \refsec{sec:exampl-treib-stack}, $TS$ denote the
  Treiber stack in \reffig{fig:TS}, and $AS$ denote the abstract stack
  in \reffig{fig:Abstract-TS}. Suppose {\tt s} in $\mcD$ is an instance
  of $TS$. Then the following is a possible observable trace of
  $\mcD[TS]$: \smallskip
  
  \noindent$tr \sdef  \langle (x, y, z) \mapsto (0, 0, 0), (x, y, z)
  \mapsto (0,1,0),(x, y, z) \mapsto (2,1, 0),  (x, y, z) \mapsto
  (2,1,1) \rangle
  $\smallskip 
  
  \noindent
  where $(x, y, z) \mapsto (0, 0, 0)$ is shorthand for the state $\{x
  \mapsto 0, y \mapsto 0, z\mapsto 0\}$, and we ignore stuttering.  
  Trace $tr$ is obtained by initialising as specified
  by {\tt Init}, then executing {\tt T1}, {\tt T2}, {\tt U1}, {\tt
    T3}, then {\tt U2} to completion; i.e.\ they each execute their
  operation call without interruption. It is straightforward to see
  that $tr$ can also be generated by $\mcD[AS]$, i.e., when using the
  abstract stack for {\tt s}. Thus $tr$ can be accepted as being
  correct. Executions can, of course, be much more complicated than
  $tr$ --- $TS$ consists of non-atomic operations, hence,
  executions of {\tt T1}, {\tt T2} or {\tt T3} may overlap with {\tt
    U1} or {\tt U2}.  \hfill\qed
\end{example}

We say that
$TS$ \emph{contextually trace refines} $AS$ \emph{with respect to the
  client program $\mcC$} iff every trace of $\mcC[TS]$ is a possible
trace of $\mcC[AS]$. In this paper, we wish to know whether contextual
refinement holds for every client program. To this end, we say $TS$
\emph{contextually trace refines} $AS$ iff $TS$ contextually trace
refines $AS$ with respect to every client program
$\mcC$.

\vspace{-1mm}

\subsection{Correctness conditions on concurrent objects}

% This paper aims to understand how correctness conditions on concurrent
% objects affect contextual trace refinement. 
There are many notions of
correctness for concurrent objects, and these are defined in terms of
\emph{histories} of invocation and response events corresponding to
operation calls on the object \cite{HeSh08}.

Concurrent histories may consist of both overlapping and
non-overlapping operation calls, inducing a partial order on
events. Safety properties define how, if at all, this partial order is
preserved by the corresponding abstract histories
generated by the corresponding sequential object
\cite{HeSh08,DDGS15-ECOOP}.  We will consider three different safety
properties. \emph{Sequential consistency} is a simple condition
requiring the order of operation calls in a concrete history for a
single process to be preserved. Operation calls performed by different
processes may be reordered in the abstract history even if the
operation calls do not overlap in the concrete history.
\emph{Linearizability} strengthens sequential consistency by requiring
the order of non-overlapping operations to be preserved. Operation
calls that overlap in the concrete history may be reordered when
mapping to an abstract history. \emph{Quiescent consistency} is weaker
than linearizability, but is incomparable to sequential consistency. A
concurrent object is said to be quiescent at some point in its history
if none of its operations are executing at that point. Quiescent
consistency requires the order of operation calls that are separated
by a quiescent point to be preserved.  Operation calls that are not
separated by a quiescent point may be reordered, including operations
performed by the same process.

Progress conditions on concurrent objects are necessary to ensure that
clients will eventually be able to continue execution after calling
operations on the objects they use. We consider a notion of progress
called \emph{minimal progress} \cite{HS11}, which guarantees that
after some finite number of steps, some operation of the concurrent
object terminates.
\vspace{-2mm}

\section{Modelling client-object systems}
\label{sec:modell-client-object-2}
Our formal framework for reasoning about contextual trace refinement
is based on existing work on action systems with procedures
\cite{SereW00}, extended to cope with potentially non-atomic
operations. We let $Var$ and $Val$ denote the types of variables and
values, respectively. A \emph{state} is a function
$\Sigma_{V} \sdef V \fun Val$, where $V \subseteq Var$, and a
\emph{predicate} of type $K$ is of type $\mcP K \sdef K \fun \bool$,
e.g., a \emph{state predicate} over $V$ is of type $\mcP \Sigma_{V}$.

The abstract syntax of an action system is of the form:\smallskip

\hfill $\begin{array}[t]{rcl} \msA & \ddef & \lact
  \begin{array}[t]{@{}l@{}}
    \lvar L ; \gvar G ; 
    \proc ph_1 = P_1 \dots \proc ph_n = P_n  ; I ;
    \ddo\ A\ \ood  \ract 
  \end{array}
\end{array}$\hfill{}\smallskip

\noindent
where $L \subseteq Var$ is a set of \emph{unobservable variables} and
$G \subseteq Var$ a set of \emph{observable variables} such that
$L \cap G = \emptyset$; each $ph_i = P_i$ is a (non-recursive)
procedure declaration; $I$ is an action modelling initialisation; and
$A$ is the main action. Within each $ph_i = P_i$, $P_i$ is an action
and $ph_i$ is a procedure heading $p_i(\val v, \res x)$ with procedure
name $p_i$ and optional call-by-value and call-by-result parameters
$v$ and $x$. Procedure declarations may additionally be parameterised
by thread identifiers.  

The abstract syntax of \emph{actions} is of the form:\smallskip

\noindent\hfill
$\begin{array}{rcl} A \ddef % [r]
  \var x \mid \rav x \mid \sskip \mid x \nasgn E \mid x \asgn e \mid
  p(e,x) \mid
  A_1 ; A_2 \mid b \to A \mid A_1 \sqcap A_2
\end{array}$\hfill\smallskip

\noindent where $x$ is a variable, $E$ is a set-valued expression, $e$
is an expression, $p$ is a procedure name and $b$ is a predicate.  Actions
$\var x$ and 
$\rav x$ introduce and remove variable $x$ from the state space,
respectively, $\sskip$ is an action that leaves the state unchanged,
$x \nasgn E$ denotes non-deterministic assignment, $x \asgn e$ denotes
assignment, $p(e,x)$ is a procedure call with value parameter $e$ and
result parameter $x$,
$A_1 ; A_2$ is sequential composition of $A_1$ and $A_2$,
$b\to A$ is a guarded action, and $A_1 \sqcap A_2$ is (demonic) choice
between $A_1$ and $A_2$. 
% $\begin{array}[t]{ll}
%      & \qquad\qquad 
% \end{array}$\hfill{}\smallskip

The meaning of \emph{parameterless procedures} is given by
syntactically replacing each procedure call $p$ in $A$ by the
procedure body, $P$.  Procedure parameters are handled by introducing
new local variables with the same name; for call-by-value, the new
variable is initialised with the value of the actual parameter, while
for call-by-results, the final value is copied to the variable passed
as the parameter.  This can be seen in Examples \ref{ex:c-as} and
\ref{ex:param} below.

%To cope
%with procedures with parameters, we adapt Sere and Walden's parameter
%passing mechanisms call-by-value $p(\val v)$ and call-by-result
%$p(\res v)$. Because procedures are potentially non-atomic, inputs and
%outputs cannot be handled in a generic manner as done in
%\cite{SereW00}. Instead, they are specific to the procedure being
%defined. Two examples of parameter passing are given in Examples
%\ref{ex:c-as} and \ref{ex:param} below.

\begin{example} 
  \label{ex:c-as}
  Consider again the client program $\mcD$ from
  \refsec{sec:exampl-treib-stack} and suppose it uses the abstract
  stack object $AS$ in \reffig{fig:Abstract-TS}. The action system
  modelling the client-object system is $\mcD[AS]$, which is given
  below. The shared stack is a sequence modelled by an unobservable
  variable $S$. The client uses variables $x$, $y$ and $z$, and
  unobservable program counters $pc_1$ and $pc_2$ are used to model
  control flow. Note that all client variables are assumed to be
  observable, but none of the object's varaibles are observable. We
  assume $npc_t(k)$ is an action that sets $pc_t$ to $k$ when the
  procedure being called has terminated. There are many possible ways
  to detect termination of the procedure being called, e.g., by
  checking whether a program counter for the procedure has been
  declared in the current state (see \refex{ex:param} below).
  \smallskip
  
  \noindent \hfill
  $ \lact \begin{array}[t]{@{}l@{}} \lvar S, pc_1, pc_2 ; \gvar x, y,
    z ;
            \\
    \begin{array}[t]{@{}l@{}}
      \proc push_t(\val in)  =  S \asgn \langle in
      \rangle \cat S \\
      \proc pop_t(\res out)  =
      \begin{array}[t]{@{}ll@{}}
        & S = \emptyseq \land \neg dec(ret) \to \var ret; ret \asgn empty  \\
        \sqcap & S \neq \emptyseq \land \neg dec(ret) \to \var ret;
                 ret, S \asgn head.S, tail.S\\ 
        \sqcap & dec(ret) \to out \asgn ret ; \rav ret\,;
      \end{array} 
    \end{array} \\
    S, pc_1, pc_2 \asgn \emptyseq, T1, U1 ; x, y, z \asgn 0, 0,
    0 ; 
    \\
    \ddo
    \begin{array}[t]{@{}l@{~}l@{}l@{}}
      & pc_1 = T1 \to  push_1(1) ; npc_1(T2) & \ \ \ \ \sqcap \ \ \ \   pc_2 = U1 \to  pop_2(y) ; npc_2(U2)
      \\
      \sqcap &  pc_1 = T2 \to  push_1(2) ; npc_1(T3)
      & 
      \ \ \ \ \sqcap \ \ \ \   pc_2 = U2 \to  z, pc_2 \asgn x, \bot  
      \\
      \sqcap & pc_1 = T3 \to  pop_1(x)  ; npc_1(\bot) & \hfill \ood \ract
    \end{array}
  \end{array}$\hfill \medskip
  % \noindent  The procedures for the push and pop operations use input
  % $npc$ to set the next $\widehat{pc}$ value for the client thread and
  % modify $S$ in the obvious way. The initialisation and the main
  % actions are also straightforward.  
  
  \noindent 
  Note that because
  $(A_1 \sqcap A_2) ; A = (A_1 ; A) \sqcap (A_2 ; A)$ holds, and
  $b_1 \to ( b_2 \to S) = b_1 \land b_2 \to S$ and
  $b \to (A_1 \sqcap A_2) = (b \to A_1) \sqcap (b \to A_2)$,
  expanding the action corresponding to $U1$ results in the action
  \[
  \begin{array}[t]{@{}ll@{}}
    & pc_2 = U1 \land S = \emptyseq \to out \asgn empty ; npc_2(U2)\\
    \sqcap & pc_2 = U1 \land S \neq \emptyseq \to out, S \asgn head.S,
    tail.S \,; npc_2(U2)
  \end{array}
\]
  \endbox
\end{example}

\begin{example}
  \label{ex:param}
  The $push_t$ operation of the Treiber stack (invoked by thread $t$)
  is defined as follows, where
  $dec(v) \sdef \lambda \sigma \dot v \in \dom \sigma$ holds iff $v$
  is declared in the domain of the given state and
  $newNode.n \sdef n \nasgn Nodes\ ; Nodes \asgn Nodes \backslash
  \{n\}$
  assigns $n$ to be a new node from the available set of nodes
  $Nodes$. For simplicity, we assume $Nodes$ is an infinite set (e.g.,
  the natural numbers), so a new node is always available. Further
  note that the object's program counter is $\hpctvar$, which is
  distinguished from the client's program counter $pc_t$. Thus we
  have: \smallskip
  
  \noindent\hfill$\begin{array}{ll}
    \proc push_t(\val in)  & {} =
    \begin{array}[t]{@{}llll@{}}
      & \neg\, dec(\hpctvar) & \to &  \var \hpctvar, v_t, n_t, ss_t ; v_t\asgn in \\
      \sqcap & \hpctvar = H1 & \to &  newNode.n_t ; \hpctvar \asgn H2\\
      % \sqcap & \hpctvar = H2 &\to &  val.n_t \asgn v_t; \hpctvar \asgn H3 \\
      & ...\\
      \sqcap & \hpctvar = H6 &\to &  \rav \hpctvar, v_t, n_t, ss_t 
    \end{array}
  \end{array}$\hfill\smallskip 

  \noindent The $pop$ operation is similar, except that it
  additionally sets the output variable to the returned
  value. \smallskip

  \noindent\hfill$
  \begin{array}[t]{ll}
    \proc pop_t (\res out)
    & {} = 
    \begin{array}[t]{@{}llll@{}}
      & \neg\, dec(\hpctvar) & \to &  \var \hpctvar, ss_t, ssn_t, lv_t \\
      & ...\\
      \sqcap & \hpctvar = P7 &\to & out \asgn 
      lv_t;  \rav \hpctvar, ss_t, ssn_t, lv_t 
    \end{array}
  \end{array}$\hfill\smallskip 

  \noindent The action system resulting from using the Treiber stack
  (which we will refer to as $TS$) as the shared concurrent object in
  \refsec{sec:exampl-treib-stack} is $\mcD[TS]$. It is similar to the
  action system in \refex{ex:c-as}, except that the unobservable
  variables are $Nodes$ (the set of all available nodes), $Head$ (a
  pointer to a node, or $null$), $val$ (a partial function of type
  $Nodes \pfun Val$), $next$ (a partial function of type
  $Nodes \fun Node$); the procedure declarations above are used; and
  initialisation of the object is
  $Nodes, Head, val, next \asgn \nat, null, \emptyset, \emptyset$.
  \endbox
\end{example}

We now make the concept of an object and the notation $\mcC[O]$ for an
object $O$ and client $\mcC$ more precise. An \emph{object} is a
triple
$O \sdef (L, \{ph_{1,t} = P_{1,t}, \dots, ph_{n,t} = P_{n,t}\}, I)$,
where $L$ is a set of variables,
$\{ph_{1,t} = P_{1,t}, \dots, ph_{n,t} = P_{n,t}\}$ is a set of
(potentially parameterised) procedure declarations, and $I$ is an
initialisation action. A \emph{client} is a triple
$\mcC \sdef (G, A, J)$, where $G$ is a set of variables, and $A$ and
$I$ are the main and initialisation actions, respectively. Then
$\mcC[O]$ is the action system\smallskip

\hfill $ \lact \lvar L ; \gvar G ; \proc ph_{1,t} = P_{1,t} \dots
\proc ph_{n,t} = P_{n,t}; I ; J ;
\ddo\ A\ \ood \ract $.\hfill{}
\vspace{-2mm}

\section{Contextual trace refinement}
\label{sec:modell-client-object}

We now give the semantics for action systems and define contextual
trace refinement, which extends the existing theory on trace
refinement \cite{back94trace}. Note that we only use part of the
action systems framework, foregoing generality in favour of a subset
of the theory adequate for handling with contextual trace
refinement. In particular, to develop a more direct link to trace
refinement, we only give a relational semantics for actions, as
opposed to the usual predicate transformer semantics.

 % The basic elements of
% our programming model are predicates and relations which we model as
% functions to booleans.  In particular, for types $K$ and $K'$, a
% \emph{predicate} is of type $\mcP K \sdef K \fun \bool$ and a
We assume expressions are functions from states to values.  A
\emph{relation} is of type $\mcR(K, K') \sdef K \fun \mcP K'$, thus a
\emph{state relation} is of type $\mcR (\Sigma_{V}, \Sigma_{V'})$,
where $V, V' \subseteq Var$. Assume $r$, $r_1$ and $r_2$ are state
relations, $b$ is a predicate and $S$ is a set.  We let:
\begin{itemize}
\item $(r_1 \circ r_2).\gamma.\gamma' \sdef \exists \gamma'' \dot
  r_1.\gamma.\gamma'' \land r_2.\gamma''.\gamma'$ 
denote \emph{relational composition},
\item 
$(b \dres r).\gamma.\gamma' \sdef b.\gamma \land r.\gamma.\gamma'$
 denote \emph{domain restriction}, and
\item $S \ndres r = \{(\gamma, \gamma') \in r \mid \gamma \notin
  S\}$ denote domain anti-restriction.
\end{itemize}
For a function $f$, we let
$f \oplus\{x \mapsto v\} \sdef \lambda z \in \dom f \dot {\bf if}\ z =
x\ {\bf then}\ v\ {\bf else}\ f.z$
denote \emph{functional overriding}.
% For predicates $p_1,
% p_2 \in \mcP K$ and relations $r_1, r_2 \in \mcR(K, K')$, we define
% \emph{universal implication} as
% \begin{align*}
%   p_1 \entails p_2 \ \ \defs \ \ \forall k : K \spot p_1.k \imp p_2.k
%   & \qquad & r_1 \entails r_2 \ \ \defs \ \ \forall k : K, k' : K' \spot
%   r_1.k.k' \imp r_2.k.k'
% \end{align*}
\begin{definition}
  The (relational) semantics of an action $A$ is given by
  $rel.A$:\smallskip

  $\begin{array}[t]{rl@{\qquad}rl}
    rel.(\var x) \sdef & 
    \lambda \sigma \dot \lambda \sigma' \dot (\{x\} \ndres \sigma') =
                         \sigma \land dec(x).\sigma'

    & rel.\sskip & \sdef  \id \\
    rel.(\rav x)  \sdef & \lambda \sigma \dot \lambda \sigma' \dot
    (\{x\} \ndres \sigma) = \sigma'
    & rel.(b \to A_1) & \sdef b \dres rel.A_1 \\
    rel.(x \asgn e)  \sdef & \lambda \sigma \dot \lambda \sigma' \dot
    \sigma' = \sigma \oplus\{x \mapsto e.\sigma\} 
 & rel.(A_1 ;
    A_2) &  \sdef rel.A_1 \circ rel.A_2\\

     rel.(x \nasgn E) \sdef& \lambda \sigma
                             \dot \lambda \sigma' \dot 
    \exists k : E.\sigma \dot \sigma' = \sigma \oplus\{x \mapsto k\}
& rel.(A_1 \sqcap
    A_2) & \sdef rel.A_1 \lor rel.A_2 \\
  \end{array}$
\end{definition}

% Note that because $\circ$ distributes over $\lor$, both
% $A ; (A_1 \sqcap A_2) = (A ; A_1 \sqcap A ; A_2)$ and
% $(A_1 \sqcap A_2) ; A = (A_1 ; A \sqcap A_2 ; A)$ hold for any actions
% $A$, $A_1$ and $A_2$. 
Recall that the semantics of a procedure call is given by
substitution as described in \refsec{sec:modell-client-object-2}.

We let $grd.A.\gamma \sdef \gamma \in \dom (rel.A)$ denote the
\emph{guard} of $A$.  Because an action system is a loop with a
non-deterministic choice over actions, we frequently use iteration in
our reasoning. Formally, finite iteration of relation $r$ (denoted
$r^*$) is defined as follows: \smallskip

\hfill
$r^0\ \sdef\ \id \qquad\qquad r^{k+1}\ \sdef \ r \circ r^k \ \
\qquad\qquad \ \ r^* \ \sdef \ \exists k \in \nat \dot
r^k$\hfill\smallskip

\noindent The semantics of an iterated action is defined by lifting
from iteration defined on relations, namely,
$rel.A^* \sdef (rel.A)^*$.  We say an \emph{iterated execution of $A$
  terminates from state $\gamma$} iff
$term.A.\gamma \sdef \exists k \dot \all \gamma' \dot
(rel.A)^k.\gamma.\gamma' \imp \neg grd.A.\gamma'$.
Note that $\neg grd.A.\gamma \imp term.A.\gamma$ holds for all actions
$A$ and states $\gamma$.

We use $\seq X$ to denote (possibly infinite) sequences of elements of
type $X$, and assume indices start from $0$.
\begin{definition}\sloppypar
  \label{def:act-sys-beh}
  A possibly infinite sequence of states $s$ is a \emph{trace} of
  action system $\msA$ iff $\exists \sigma \dot rel.I.\sigma.(s.0)
  \land \all i: \dom s \bs \{0\} \dot rel.A.(s.(i - 1)).(s.i)$
  holds.\endbox
\end{definition}\noindent
A \emph{trace} is \emph{complete} iff either the trace is of infinite
length or the guard of $A$ does not hold in the last state of the
trace. The set of all \emph{complete traces} of an action system
$\msA$ is denoted $\beh \msA$.

Traces (\refdef{def:act-sys-beh}) provide a conceptually simple model
for a system's execution, and trace refinement provides a conceptually
simple notion of substitutability \cite{back94trace}. Typically,
because a concrete system is more fine-grained than the abstract, one
must remove \emph{stuttering} from a trace, i.e., consecutive states
that leave the observable state unchanged. An action system may also
exhibit \emph{infinite stuttering} by generating a trace that ends
with an infinite sequence of consecutive stuttering steps. After
infinite stuttering, one will never be able to observe any state
changes, and hence, we treat infinite stuttering as \emph{divergence},
which is denoted by a special symbol `$\diverge \notin \Sigma$'.  For
any trace $s \in \llbracket \msA \rrbracket $, we define $Tr.s$ to be
the non-stuttering observable sequence of states, possibly followed by
$\diverge$, which is obtained from $s$ as follows. First, we obtain a
sequence $s'$ by removing all finite stuttering in $s$ and replacing
any infinite stuttering in $s$ by $\diverge$. Second, for each
$i \in \dom s'$, we let
$(Tr.s).i = {\bf if}\ s'.i \neq \diverge\ {\bf then}\ L \ndres s'.i\
{\bf else}\ \diverge$.
It is straightforward to define functions that formalise the steps and
above (see for example \cite{Don09}).
\begin{definition}
  We say abstract action system $\msA$ is \emph{trace refined} by
  concrete action system $\msC$ (denoted $\msA \sref \msC$) iff $\all
  s' \in \beh{\msC} \dot \exists s \in \beh{\msA} \dot Tr.s = Tr.s'$
  holds.\endbox
\end{definition}
Back and von Wright have developed simulation rules (details elided
due to lack of space) for verifying trace refinement of action systems
\cite{back94trace}, which we adapt to reason about client-object
systems in Lemmas \ref{lem:co-fsim} and \ref{lem:co-bsim}. First, we
formalise the meaning of contextual trace refinement.  The notion is
similar to the notion of data refinement given by He et al.\
\cite{HeHS86,deRoever98}, but extended to traces, which enables one to
cope with non-terminating reactive systems.
\begin{definition}
  An abstract object $OA$ is \emph{contextually trace refined} by a
  concrete object $OC$, denoted $OA \srefc OC$, iff for any client
  $\mcC$ we have $\mcC[OA] \sref \mcC[OC]$.\endbox
\end{definition}

% To simplify reasoning about trace refinement, states are often split
% into \emph{unobservable} and \emph{observable} parts by partitioning
% $V$ into $L$ and $G$, representing the unobservable and observable
% variables, respectively, where $V = L \cup G$ and $L \cap G =
% \emptyset$. We define notation $\Sigma_{V} \sdef V \fun Val$ and
% $\Sigma_{L,G} \sdef \Sigma_L \times \Sigma_G$. 

In this paper, for simplicity, we assume that (atomic) actions do not
abort \cite{deRoever98}, therefore the proof obligations for aborting
actions do not appear in Lemmas \ref{lem:co-fsim} and
\ref{lem:co-bsim} below -- it is straightforward to extend our results
to take aborting behaviour into account. However, like Back and von
Wright \cite{back94trace}, our notion of refinement ensures
\emph{total correctness} of the systems we develop, i.e., the concrete
system may only deadlock (or diverge) if the abstract system deadlocks
(or diverges). Thus, in addition to the standard step correspondence
proof obligations for ensuring safety of the concrete system, we
include Back and von Wright's proof obligations that ensure progress.

% \reflem{lem:co-fsim} below gives forward simulation rules for
% \emph{contextual trace refinement}; it is straightforward to show that
% these imply Back and von Wright's proof obligations
% \cite{back94trace}. Conditions \refeq{eq:l9} and \refeq{eq:l10-s}
% guarantee safety, whereas \refeq{eq:l11} and \refeq{eq:l12} guarantee
% progress. 
Because the entire state of the client is observable, the proof
obligations pertaining to the client can be trivially discharged,
leaving one with proof obligations that only refer to the object.  For
procedure declarations
% $\{p1_t(\val v, \res x) = P1_t, \dots, pN_t(\val v, \res x) = PN_t\}$,
$P \sdef \{ph_{1,t} = P_{1,t}, \dots, ph_{n,t} = P_{n,t}\}$, we let 
\[
\begin{array}[t]{@{}ll@{}}
  act.P \sdef \textstyle\bigsqcap_{v, x, t}\
  p_{1,t}(v,x) \sqcap \dots \sqcap p_{n,t}(v,x)
\end{array}
\]
denote the action corresponding to the potential procedure calls in
$P$, then define the following action, where $tt_t$ is assumed to be a
fresh variable for all threads $t$.
\[
rem.P \sdef \textstyle\bigsqcap_{v, x, t}\ % \neg tt_t
\neg dec(\hpctvar) \to
\begin{array}[t]{@{}l@{}}
  (p_{1,t}(v,x) \sqcap \dots \sqcap p_{n,t}(v,x))
  % (\neg dec(\hpctvar)
  % \to tt_t \asgn true \sqcap dec(\hpctvar) \to \sskip)
\end{array}
\]

The guard $\neg dec(\hpctvar)$ is used to detect whether the procedure
being executed by thread $t$ has terminated. Upon termination of
procedure $p_{i,t}$ for some $1 \le i \le n$, $\neg dec(\hpctvar)$
will hold. The intention is to use $rem.P$ in \refeqn{eq:l12} below,
which attempts to execute the remaining steps of the operation invoked
by thread $t$ to completion.

% Further define
% $unblock.\tau.\tau'.\sdef $ as the relation that sets each $tt_t$ in
% $\tau$ to $false$.

\begin{lemma}[Forward simulation] 
  \label{lem:co-fsim}
  Suppose $OA = (L_A, P_A, I_A)$ and $OC = (L_C, P_C, I_C)$ are
  objects. Then $OA \srefc OC$ if there exists a relation $R$ and the
  following hold for any states $\sigma$, $\tau$ and $\tau'$:
 % an
  % $R \in \mcR(\Sigma_{L_A}, \Sigma_{L_C})$ and the following hold for
  % any states $\sigma % \in \Sigma_{L_A}
  % $,
  % $\tau, \tau' \in \Sigma_{L_C}$:
  \vspace{-2mm}
  \begin{eqnarray}
    \label{eq:l9}
    rel.I_C.\tau' & \imp & \exists \sigma \dot
    R.\sigma.\tau'  \land
    rel.I_A.\sigma
    \\
    \label{eq:l10-s}
    R.\sigma.\tau \land rel.(act.P_C).\tau.\tau' &
    \imp & \exists \sigma' \dot R.\sigma'.\tau' \land
    rel.(act.P_A)^*.\sigma.\sigma'
    % \\
    % R.\sigma.\tau \land abort.OC.\tau & \imp & abort.OA.\sigma
    \\
    \label{eq:l11} 
    R.\sigma.\tau \land \neg grd.(act.P_C).\tau & \imp & \neg
    grd.(act.P_A).\sigma
    \\
    \label{eq:l12}
    \begin{array}[b]{@{}r@{}}
      % R.\sigma.\tau \land
      \tau' = (\tau \oplus \textstyle \bigcup_{t:T}
      \{tt_t \mapsto false\})
    \end{array}
                  & \imp & term.(rem.P_C).\tau'  % \neg term.(rem.P_A).\sigma'
  \end{eqnarray}
\end{lemma}
\noindent The first three proof obligations are straightforward. Proof
obligation \refeq{eq:l12} requires that the main action of the
concrete object $OC$ terminates if threads do not invoke new
operations after the operation currently being executed has
terminated. Note that \refeqn{eq:l12} does not rule out infinite
stuttering within the program $\mcC[OC]$, but it does ensure that any
infinite stuttering is caused by the client as opposed to the object
$OC$, and hence, this infinite stuttering must also be present within
$\mcC[OA]$. Therefore, if \refeq{eq:l12} holds, so does Back and von
Wright's non-termination condition.

% Forward simulation is sound, but not complete for verifying
% trace refinement \cite{back94trace,deRoever98}. Therefore, d
Dually to forward simulation, there exists a method of \emph{backward
  simulation}, which requires that the abstract action system under
consideration is \emph{continuous}. An action system $\msA$ with main
action $A$ is \emph{continuous} iff for all $\sigma$, the set
$\{\sigma' \mid rel.A.\sigma.\sigma'\}$ is finite, i.e., $A$ does not
exhibit infinite non-determinism.
\begin{lemma}[Backward simulation]
  \label{lem:co-bsim}
  Suppose $OA = (L_A, P_A, I_A)$ and $OC = (L_C, P_C, I_C)$ are
  objects and $\mcC$ is a client such that $\mcC[OA]$ is
  continuous. Then $\mcC[OA] \sref \mcC[OC]$ holds if there exists a total relation
  $R % \in \mcR(\Sigma_{L_A}, \Sigma_{L_C})
  $ and for any states 
  $\sigma' % \in \Sigma_{L_A}
  $ and
  $\tau, \tau' % \in \Sigma_{L_C}
  $ \refeq{eq:l12}
  holds and each of the following hold:\vspace{-2mm}
  \begin{eqnarray}
    \label{eq:1}
    rel.I_C.\tau' \land R.\sigma'.\tau' & \imp & rel.I_A.\sigma'
    \\
    \label{eq:2}
    rel.(act.P_C).\tau.\tau' \land R.\sigma'.\tau'  & \imp & \exists 
    \sigma \dot R.\sigma.\tau \land rel.(act.P_A)^*.\sigma.\sigma'
    \\
    \label{eq:13}
    \neg grd.(act.P_C).\tau & \imp & \exists \sigma \dot R.\sigma.\tau \land \neg
    grd.(act.P_A).\sigma
    % \\
    % \label{eq:3}
    % \neg term.(rem.P_C).\tau & \imp & \exists \sigma \dot R.\sigma.\tau \land \neg
    %                                   term.(rem.P_A).\sigma
  \end{eqnarray}
\end{lemma}
Lemmas \ref{lem:co-fsim} and \ref{lem:co-bsim} reduce the proof
obligations for trace refinement of client-object systems to the level
of objects only. This provides us with the opportunity to explore
properties of objects in isolation to guarantee contextual trace
refinement.

\vspace{-2mm}

\section{Events and histories}

% Thus far, we have formalised the syntax and semantics of client-object
% systems, and given the meaning of contextual trace
% refinement. % Recall, that our ultimate aim is to understand the link
% between different notions of safety
This section provides background for defining safety (e.g.,
linearizability) and progress (e.g., lock-freedom) properties of
concurrent objects \cite{HeSh08}. We define both types of properties
in terms of \emph{histories} of invocation and response events
\cite{Herlihy90,HeSh08} that record the externally visible interaction
between a client and the object it uses. The type of an event is
$Event$, which is defined as follows~\cite{DSW11TOPLAS}:\smallskip

\hfill 
$Event \ \  ::= \ \  inv \lang \nat \times Op \times (Val \cup \{\bot\}) \rang \mid ret \lang \nat
\times Op \times (Val \cup \{\bot\}) \rang$\hfill{} \smallskip

\noindent
The components of each event are the thread identifier, the operation
name and input/output values. We use $\bot \notin Val$  to
denote an invocation (return) event that has no input (output). Thus,
for example, $inv(1, push, 2)$ denotes an $push$ invocation by thread
$1$ with value $2$, and $ret(1, push, \bot)$ denotes a return from
this invocation.  

The history of an object is a (potentially infinite) sequence of
events, i.e., $History \sdef \seq Event$. A history of an object is
generated by an execution of a \emph{most-general client} for the
object \cite{Doherty03}. We formalise the concept of a most general
client in our framework in \refdef{def:mgc} below, but first we
describe how invocations and responses are recorded in a history. For
an object
$O \sdef (L, \{ph_{1,t} = P_{1,t}, \dots, ph_{n,t} = P_{n,t}\}, I)$
assuming $H \notin L$ is a history variable, we let $P_{i,t}^H$ be the
\emph{history-extended} action derived from $P_{i,t}$ by additionally
recording invocation and response events in $H$ (also see~\cite{DSW11TOPLAS}).
\begin{example}
\label{ex:hist-ext}
The history-extended action for $push_t$ from \refex{ex:c-as}
is:\smallskip

\hfill $H \asgn H \cat \langle inv(t, push, in) \rangle ; 
S \asgn \langle in
\rangle \cat S ;  H \asgn H \cat \langle ret(t, push, \bot) \rangle$\hfill \smallskip

\noindent while the history-extended version of $push_t$ procedure
from \refex{ex:param} is: 
\smallskip

\hfill
$\begin{array}[b]{@{}llll@{}}
    & \neg\, dec(\hpctvar) & \to &  \var \hpctvar, v_t, n_t, ss_t ; v_t\asgn
    in ;   H \asgn H \cat \langle inv(t, push, in) \rangle  \\
    % \sqcap & \hpctvar = H1 & \to &  newNode.n_t ; \hpctvar \asgn H2\\
    % \sqcap & \hpctvar = H2 &\to &  val.n_t \asgn v_t; \hpctvar \asgn H3 \\
    & ...\\
    \sqcap & \hpctvar = H6 &\to &  H \asgn H \cat \langle ret(t, push,
    \bot) \rangle ;  \rav \hpctvar, v_t, n_t, ss_t 
\end{array}$\hfill \endbox

\end{example}

\begin{definition}
  \label{def:mgc}
  The \emph{most general client} of $O$ is the action system below,
  where $H \notin L$ is its history and $tt \notin L$ is a fresh
  variable that models termination.  \smallskip

  \hfill$ \MGC[O] \sdef 
  \begin{array}[t]{@{}l@{}}
    \lact 
    \begin{array}[t]{@{}ll@{}} \lvar L \cup \{H, tt\} ; \gvar
      \emptyset ; \\
      % \proc p1_t(\val v, \res x) = P1_t^H\ \dots \proc
      % pN_t(\val v, \res x) = PN_t^H
      \proc ph_{1,t} = P_{1,t}^H\  \dots \proc ph_{n,t} =
      P_{n,t}^H\ ;
      \\
      I ; H, tt \asgn \emptyseq, false\ ; \\
      \ddo\ \neg tt
      \to \textstyle\bigsqcap_{v, x, t}\ p1_t(v, x) \sqcap \dots \sqcap pN_t(v, x)
      \sqcap tt \asgn true\ \ood \ract
    \end{array}
  \end{array}
  $\hfill{\ }
\end{definition}

\noindent Thus, $\MGC[O]$ includes unobservable variables $H$
(initially $\emptyseq$) and $tt$ (initially $false$), which model the
history and termination of $\MGC[O]$, respectively. Provided $tt$ is
false, the history-extended procedures of $O$ are executed, or the
system decides to terminate by setting $tt$ to $true$. The intention
of $\MGC[O]$ is to model all possible client behaviours, including for
instance faults (where a thread stops running) or a divergence (where
a thread repeatedly executes the same operation).

\begin{definition}
  The set of histories of an object $O$ is given by\smallskip

\noindent\hfill  $\{h \in \seq Event | \exists s : \llbracket \MGC[O] \rrbracket \dot
  \exists i : \dom s \dot h = (s.i).H\}$. \hfill{\ }
\end{definition}

% The formal definition of linearizability relies on a number of
% preliminary definitions \cite{DSW11TOPLAS}.  For a history $h$ and
% thread $t$, we let $h_{|t}$ denote the maximal subsequence of $h$ such
% that for each event $e \in h_{|t}$, $\pi_1.e = t$

  \vspace{-4mm}

% sense that a partially executed action cannot be interleaved with
% another action, or in other words, interleaving only occurs at the
% level of actions.

\section{Contextual trace refinement: Progress}
\label{sec:corr-cond}

The progress condition we will consider is \emph{minimal progress},
which guarantees system-wide progress, even though there may be
individual threads that may not make progress \cite{HS11}. % There are
% several possible formalisations of this property, our definition is
% based on histories.
To formalise minimal progress, we say event $e_1$ \emph{matches} $e_2$
iff %$matches(e_1, e_2)$ holds where
$matches(e_1, e_2) \sdef \exists t, o, u, v \dot e_1 =
\invoke{t}{o}{u} \land e_2 =\return{t}{o}{v}$
holds, i.e., $e_1$ is an invocation of an operation by a thread and
$e_2$ is the corresponding return. We say $m \in \dom h$ is a
\emph{pending invocation} iff
$pi(m,h) \sdef \all n \in \dom h \dot m < n \imp \neg matches(h.m,
h.n)$ holds.

An object $O$ satisfies minimal progress iff for every trace $tr$ of
the $\MGC[O]$, it is always the case that in the future, either
$\MGC[O]$ terminates, or there is some pending operation invocation
that completes and returns. % Naturally, this is formalised using LTL
% operators $\Box$ (always), $\Diamond$ (sometime) and $\ocircle$
% (next), where $s \vdash F$ denotes that the LTL formula $F$ is valid
% for sequence $s$ \cite{MP92}.
\begin{definition}
  An object $O$ satisfies \emph{minimal progress} iff for every
  $s \in \llbracket \MGC[O] \rrbracket$ and $i \in \dom s$, there
  exists a $j \in \dom s$ such that $i \le j$ and
  $$(s.j).\ttvar \lor \exists m \dot pi(m, (s.j).H) \land \neg pi(m,
  (s.(j+1)).H)$$
\end{definition}
That is, for any trace $s$ of $\MGC[O]$ and index $i \in \dom s$ there
is a state $s.j$ (where $j \ge i$) from which some pending operation
in $s.j$ completes. There are a variety of objects that satisfy
minimal progress, e.g., wait-, lock-free objects under any scheduler,
and obstruction-free objects under isolating schedulers (see
\cite{HS11} for details). Objects that do not satisfy minimal progress
include obstruction free implementations that are executed using a
weakly fair scheduler.

The lemma below states that any object that satisfies minimal progress
does not suffer from deadlock, and is guaranteed to terminate if no
additional operations are invoked.
\begin{lemma}
  \label{lem:lock-free}
  If $O = (L, P, I)$ satisfies minimal progress, then for any
  $\gamma \in \llbracket \MGC[O] \rrbracket$ and $i \in \dom \gamma$,
  both $grd.(act.P).(\gamma.i)$ and
  condition % $term.(rest.P).(\gamma.i)$
  \refeqn{eq:l12} hold.
\end{lemma}
Using \reflem{lem:lock-free}, we simplify and combine Lemmas
\ref{lem:co-fsim} and \ref{lem:co-bsim}. In particular, we are left
with %  to
% discharge the blocking conditions, \refeqn{eq:l11} and
% \refeqn{eq:13}, and termination condition
% \refeqn{eq:l12} % and \refeqn{eq:3},
% leaving one with
the proof obligations for safety only as in the theorem below.
\begin{theorem}
  \label{thm:co-safety}  
  Suppose $OA = (L_A, P_A, I_A)$ and $OC = (L_C, P_C, I_C)$ are
  objects, $OC$ satisfies minimal progress, and $R \in
  \mcR(\Sigma_{L_A}, \Sigma_{L_C})$. Then\vspace{-2mm}
  \begin{enumerate}
  \item $OA \srefc OC$ if both \refeqn{eq:l9} and \refeqn{eq:l10-s}
    hold, and
  \item for any client $\mcC$ such that $\mcC[OA]$ is continuous,
    $\mcC[OA] \sref \mcC[OC]$ holds if $R$ is total and both
    \refeqn{eq:1} and \refeqn{eq:2} hold.
  \end{enumerate}
\end{theorem}
\vspace{-2mm}

\section{Safety and contextual trace refinement}

We give the formal definition of safety properties using the
nomenclature in \cite{DDGS15-ECOOP} and \cite{DSW11TOPLAS}. We say
$m, n \in \dom h$ form a \emph{matching pair} in $h$ iff $mp(m,n,h)$
holds, where $ mp(m,n,h) \sdef
  m < n \land matches(h.m, h.n)\land 
  \all i 
  \dot m < i < n\imp  \pi_1.(h.i) \neq \pi_1.(h.m)
$
and $\pi_i$ is the \emph{projection function} returning the $i$th
element of the given tuple.

Following \cite{DDGS15-ECOOP}, safety properties are defined in terms
of a history $h$ and a \emph{mapping function} $f$ between
indices. The \emph{sequential history} corresponding to $h$ and $f$ is
obtained using
$map(h,f) \sdef \{f(k) \mapsto h(k) \mid k \in \dom f\}$.  Different
safety properties are defined by placing different types of
restrictions on $f$. The most basic restriction is validity of a
mapping.  We say a function $f$ is a \emph{valid mapping function} if,
for any history $h$, (a) the domain of $f$ is contained in the domain
of $h$, (b) the range of $f$ is a consecutive sequence starting from
$0$, (c) $f$ only maps matching pairs in $h$, and (d) matching pairs
in $h$ are mapped to consecutive events in the target abstract
history. Assuming $[m,n]$ is the set of integers from $m$ to $n$
inclusive, we formalise validity for mapping functions using
$VMF(h, f)$, where\smallskip

\hfill$  \begin{array}[t]{@{}rcl@{}}
  VMF(h, f) & \sdef & \begin{array}[t]{@{}l@{}}
    \dom f \subseteq \dom h \land  
    (\exists n : \nat \dot \ran f = [0,n-1]) %  (\all n : \dom f \dot h.n \in
    % IEvent)
    \land injective(f) \land \\
    \begin{array}[t]{@{}l@{}}
      (\all m, n : \dom h \dot
      mp(m,n,h) \imp (m \in \dom f
      \iff n \in \dom f)) \land \\
      (\all m, n : \dom f \dot
      mp(m,n,h)\imp f.n = f.m + 1)
    \end{array}
  \end{array}
  \end{array}$ \hfill{}\smallskip 

  When formalising correctness conditions, one must also consider
  \emph{incomplete histories}, which have pending operation
  invocations that may or may not have taken effect. To cope with
  these, like Herlihy and Wing \cite{Herlihy90}, we use \emph{history
    extensions}, which are constructed from a history $h$ by
  concatenating a sequence of returns corresponding to some of the
  pending invocations of $h$.
\begin{definition}
  A concurrent object $OC$ implementing an abstract object $OA$ is
  \emph{correct} with respect to a correctness condition $Z$, denoted
  $OC \models_{OA} Z$, iff for any history $h$ of $OC$, there exists
  an extension $he$ of $h$, a valid mapping function $f$ such that
  $Z(he, f)$ holds and $map(he, f)$ is a history of $OA$. \endbox
\end{definition}
% Note a history of an object is generated by its most-general client,
% and hence, is by definition \emph{well-formed}, i.e., for each thread
% $t$, history $H$ restricted to all events of $t$ is \emph{sequential}
% \cite{Herlihy90}.% (is either empty or an alternating sequence of alternating
% % matching events starting and possibly ending with an invocation.

% We say $h$ is \emph{sequential} iff it is the empty sequence or a
% sequence of alternating matching events starting and possibly ending
% with an invocation and \emph{concurrent} iff it is not
% sequential. % Because each action models an atomic transition in the

\vspace{-4mm}

\subsection{Linearizability}
\label{sec:link-corr-cont}

We now show that linearizability is a sufficient safety condition for
discharging the remaining proof obligations in \refthm{thm:co-safety}.
Linearizability is a \emph{total} condition, which means that all
completed (i.e., returned) operation calls in a given history $h$ must
be mapped by $f$.\footnote{This is in contrast to \emph{partial}
  conditions defined for relaxed memory (see \cite{DDGS15-ECOOP} for
  details).} In addition, it must satisfy an \emph{order} condition
$lin$, which states that the return of an operation may not be
reordered with an invocation that occurs after it. For an event $e$,
we use $inv?(e) \sdef \exists t, o, v \dot e = \invoke{t}{o}{v}$ to
determine whether $e$ is an invocation event and
$ret?(e) \sdef \exists t, o, v \dot e = \return{t}{o}{v}$ to determine
whether $e$ is an response event event.  \smallskip

\noindent\hfill$\begin{array}{rcl}
  total(h, f) & \sdef & \all m : \dom h \dot \neg pi(m, h) \imp m \in
  \dom f \\[1mm]
  lin(h,f) & \sdef & \all m,n : \dom f \dot m < n \land  ret?(h.m)
  \land inv?(h.n) \imp f.m < f.n
\end{array}$\hfill{}

\begin{definition}
  \label{def:intra-proc-cons}
  An object $OC$ is \emph{linearizable} with respect to $OA$ iff $OC
  \models_{OA} lin \land total$.
\end{definition}

% \subsection{Linearizability and contextual trace refinement}
% \label{sec:corr-cond}

% Our ultimate goal is to show any linearizable implementation ensures
% contextual trace refinement. 
First, we show contextual trace refinement for \emph{canonical
  implementation} \cite{Lyn96-DA,CDG05,SDW14}, which splits each
sequential abstract operation call into three actions: an
\emph{invocation}, a \emph{effect action} and a \emph{response}.
\begin{definition}
  For an abstract procedure $p_t(\val in, \res out) = P_t$, the
  \emph{canonical implementation} of the procedure is:\smallskip

  \hfill$\begin{array}{rcl}
    P^\xi_t & \sdef  &
    \begin{array}[t]{@{}l@{~}l@{~}l@{}}
      & \neg dec(\pctvar)  % \land v \neq empty
      & \to\ \var \pctvar; \pctvar \asgn 1 ; H \cat \langle inv(t, p,
        in) \rangle
      \\
      \sqcap & \pctvar = 1 & \to\ p_t(in, out) ; \pctvar \asgn 2
      \\
      \sqcap & \pctvar = 2 & \to\ \rav  \pctvar ; H \cat \langle
                             ret(t, p, out) \rangle
    \end{array}
  \end{array}$\hfill{\ }
\end{definition}

Invocation and response actions modify the auxiliary
history variable by recording the corresponding event, while the
effect action has the same effect as the abstract operation
call. Unlike the abstract object, the histories of a canonical
implementation are potentially concurrent.

\begin{theorem}[Canonical contextual trace refinement]
  \label{thm:otr-can-lin}
  Suppose $OA$ and $OB$ are objects, where $OB$ is a canonical
  implementation of $OA$. Then $OA \srefc OB$.
\end{theorem}
\noindent{\it Proof.}
We use \reflem{lem:co-fsim} because $OB$ may not satisfy minimal
progress. Here, $rel.act.OB$ trivially satisfies \refeqn{eq:l12}
because by nature each procedure of a canonical object terminates. The
proof of \refeqn{eq:l11} requires further consideration because
$rel.act.OB$ may deadlock. For example, $OB$ may be a stack with a
$pop$ operation that blocks when the stack is empty. In such cases,
because no data refinement is performed, the guard of the canonical
object is false when the guard of the abstract object is false,
allowing one to discharge \refeqn{eq:l11}. The remaining proof
obligations are straightforward. \endbox\smallskip

Next, we restate a result by Schellhorn et al.\ \cite{SDW14}, who have
shown completeness of backward simulation for verifying
linearizability. In particular, provided $OC$ is a linearizable
implementation of $OA$, they show that it is always possible to
construct a backward simulation relation between the $OC$ and the
canonical implementation of $OA$.
\begin{lemma}[Completeness of backward simulation \cite{SDW14}]
  \label{lem:exists-bsim}
  Suppose $OA, OB$ and $OC$ are objects and $\MGC[OA]$ is
  continuous. If $OC \models_{OA} lin \land total$ and $OB$ is a
  canonical implementation of $OA$, then there exists a total relation
  $R$ such that both \refeqn{eq:1} and \refeqn{eq:2} hold between
  $\MGC[OB]$ and $\MGC[OC]$.
\end{lemma}
% Note that \reflem{lem:exists-bsim} only pertains to the objects at
% hand, and does not guarantee anything to the clients using the
% objects. % Hence, for example, progress properties are not considered.

Finally, we prove our main result for linearizability, i.e., that
linearizability and minimal progress together preserves contextual
trace refinement.%  The proof uses the results we have developed along
% the way.
\begin{theorem}
  \label{thm:otr-lin}
  Suppose object $OC$ is linearizable with respect to $OA$, $OC$
  satisfies minimal progress, and $\mcM[OA]$ is continuous. If $\mcC$
  is a client such that $\mcC[OA]$ is continuous then
  $\mcC[OA] \sref \mcC[OC]$.
\end{theorem}

% \vspace{-2mm}

\noindent {\em Proof.}
Construct a canonical implementation $OB$ of $OA$. By transitivity of
$\sref$, the proof holds if both (a) $\mcC[OA] \sref \mcC[OB]$ and (b)
$\mcC[OB] \sref \mcC[OC]$.  Condition (a) holds by
\refthm{thm:otr-can-lin}, and (b) holds by \refthm{thm:co-safety}
(part 2), followed by \reflem{lem:exists-bsim}. Application of
\refthm{thm:co-safety} (part 2) is allowed because if $\mcC[OA]$ is
continuous then $\mcC[OB]$ is continuous, whereas application of
\reflem{lem:exists-bsim} is allowed because if $R$ satisfies
\refeqn{eq:1} and \refeqn{eq:2} for $\MGC[OB]$ and $\MGC[OC]$, then
$R$ also satisfies \refeqn{eq:1} and \refeqn{eq:2} for $\mcC[OB]$ and
$\mcC[OC]$.\hfill \qed

\vspace{-1mm}

\subsection{Sequential and quiescent consistency}
\label{sec:link-corr-cont}

We now consider contextual trace refinement for concurrent objects
that satisfy sequential consistency and quiescent consistency, both of
which are weaker than linearizability. Both conditions are total
\cite{DDGS15-ECOOP}. Additionally, sequential consistency disallows
reordering of operation calls within a thread (see $sc$ below), while
quiescent consistency (see $qc$ below) disallows reordering across a
quiescent point (defined by $qp$ below).\smallskip

\noindent\hfill$\begin{array}{rcl}
  sc(h, f) & \sdef & \all m, n : \dom f
  \dot \begin{array}[t]{@{}l@{}}
    m < n \land \pi_1.(h.m) = \pi_1.(h.n) \land \\
    ret?(h.m) \land inv?(h.n) \imp f.m < f.n
  \end{array} 
  \\[1mm]
  qp(m, h) & \sdef & \forall n : \dom h \dot (n \leq m \imp \neg pi(n,h[0..m]))
  \\[1mm]
  qc(h, f) & \sdef & \all m, k, n : \dom f \dot m < k < n \land 
  qp(k, h) \imp f.m < f.n
\end{array}$\hfill{}

\begin{definition}
  \label{def:intra-proc-cons}
  An object $OC$ is \emph{sequentially consistent} with respect to
  $OA$ iff $OC \models_{OA} sc \land total$, and $OC$ \emph{quiescent
    consistent} with respect to $OA$ iff $OC \models_{OA} qc \land
  total$.
\end{definition}

Our results for sequential consistency and quiescent consistency are
negative --- neither condition guarantees trace refinement of the
underlying clients, regardless of whether the client program in
question is \emph{data independent}, i.e., the state spaces of the
client threads outside the shared object are pairwise disjoint.
\begin{theorem}
  \label{thm:sequ-cons-2}
  Suppose object $OC$ is sequentially consistent with respect to
  object $OA$. Then it is not necessarily the case that $OA \srefc OC$
  holds.
\end{theorem}
\noindent{\it Proof.}
Consider the program in \reffig{fig:ce-sc}, where the client threads
are data independent --- {\tt x} is local to thread {\tt 1}, while
{\tt y} and {\tt z} are local to thread {\tt 2} --- and $s$ is assumed
to be sequentially consistent.  Suppose thread {\tt 1} is executed to
completion, and then thread {\tt 2} is executed to completion. Because
{\tt s} is sequentially consistent, the first {\tt pop} (at {\tt T3})
may set {\tt x} to {\tt 1}, the second (at {\tt U2}) may set {\tt y}
to {\tt 2}. This gives the execution:\smallskip
  
  \noindent\hfill$
  \langle (x, y, z) \mapsto (0, 0, 0), \ \ (x, y, z) \mapsto (1,0,0),
  \ \ (x, y, z) \mapsto (1,0,1), \ \ (x, y, z) \mapsto
  (1,2,1)\rangle$\hfill\smallskip
  
  \noindent which cannot be generated when using the abstract
  stack $AS$ from \reffig{fig:Abstract-TS} for {\tt s}. \hfill \qed \smallskip

\begin{figure}[t]
  \noindent
  \begin{minipage}[b]{0.46\linewidth}
    \small{\tt Init x, y, z = 0;}
    
    \begin{minipage}[t]{0.48\columnwidth}
      \small\tt Thread 1 ==
      
      T1:\ s.push(1);
      
      T2:\ s.push(2);
      
      T3:\ s.pop(x);
      
      % \ \ \ T4:\ x \asgn\ \outt;
      
    \end{minipage}
    \hfill
    \begin{minipage}[t]{0.48\columnwidth}
      \small\tt Thread 2 ==
      
      U1:\ z \asgn\ 1;
      
      U2:\ s.pop(y);
      
      % \ \ \ U3:\ y \asgn\ \outu;
    \end{minipage}
    \caption{Counter example for contextual trace refinement and
      sequential consistency}
    \label{fig:ce-sc}
  \end{minipage}
  \hfill
  \begin{minipage}[b]{0.45\linewidth}
    \small{\tt Init x, y, z = 0;}
    
    \begin{minipage}[t]{0.49\linewidth}
      \small\tt Thread 1 ==
      
      T1:\ s.push(1);
      
      % \ \ \ T2:\ z\ \asgn\ 1;
      
      T2:\ s.push(2);
      
      T3:\ s.pop(x);
      
      % \ \ \ T4:\ x \asgn\ \outt;
      
      T4:\ s.pop(y);
       
        % \ \ \ T6:\ y \asgn\ \outt;

        T5:\ s.push(3);
        
      \end{minipage}
      \hfill 
      \begin{minipage}[t]{0.45\columnwidth}
        \small\tt Thread 2 ==
        
        U1:\ s.pop(z)
        
      \end{minipage}
      \caption{Counter example for contextual trace refinement and
        quiescent consistency}
    \label{fig:ce-qc}
    \end{minipage}
    \vspace{-4mm}
\end{figure}

\refthm{thm:sequ-cons-2} differs from the results of Filipovi\'{c} et
al.\ \cite{FORY10}, who show that for data independent clients,
sequential consistency implies observational refinement. In essence,
their result holds because observational refinement only considers the
initial and final states of a client program --- the intermediate
states of a client's execution are ignored. Thus, internal reorderings
due to sequentially consistent objects have no effect when only
observing pre/post states. One can develop hiding conditions so that
observational refinement becomes a special case of contextual trace
refinement, allowing one to obtain the result by Filipovi\'{c} et al
\cite{FORY10}. Full development of this theory is left for future
work.

We now give our result for quiescent consistency.
\begin{theorem}
  \label{thm:qc-2}
  Suppose object $OC$ is quiescent consistent with respect to object
  $OA$. Then it is not necessarily the case that $OA \srefc OC$ holds.
\end{theorem}
\noindent{\it Proof.}
Consider the program \reffig{fig:ce-qc}, where the client threads are
data independent --- {\tt x} and {\tt y} are local to thread {\tt 1},
while {\tt z} is local to thread {\tt 2} --- and $s$ is a quiescent
consistent stack. The concrete program may generate the following
observable trace: \noindent\smallskip

  \noindent\hfill$
  \langle (x, y, z) \mapsto (0, 0,
  0), \ \ (x, y, z) \mapsto (1,0,0), \ \ (x, y, z) \mapsto (1,2,0), \
  \ (x, y, z) \mapsto (1,2,3)\rangle
  $\hfill\smallskip

  \noindent Note that the $pop$ operations at $T3$ and $T4$ have been
  reordered, which could happen if the execution of $pop$ at $U1$
  overlaps with $T1$, $T2$, $T3$ and $T4$. The trace above is not
  possible when the client uses the abstract stack $AS$ from
  \reffig{fig:Abstract-TS}.  \hfill \qed
\vspace{-2mm}

\section{Conclusions}

In this paper, we have developed a framework, based on action systems
with procedures, for studying the link between the correctness
conditions for concurrent objects and contextual trace refinement,
which guarantees substitutability of objects within potentially
non-terminating reactive clients. Thus, we bring together the
previously disconnected worlds of correctness for concurrent objects
and trace refinement within action systems. We have shown that
linearizability and minimal progress together ensure contextual trace
refinement, but sequential consistency and quiescent consistency are
inadequate for guaranteeing contextual trace refinement regardless of
whether clients communicate outside the concurrent object. The
sequential consistency result contrasts earlier results for
observational refinement, where sequential consistency is adequate
when clients only communicate through shared objects~\cite{FORY10}.

 % The
% current treatment refines this existing work, and presents a full
% set of results for fine-grained implementations, where both safety
% and progress properties are considered.  This problem was first
% studied formally for terminating clients \cite{FORY10}, then
% extended to potentially non-terminating clients for linearizable
% objects \cite{GY11,LiangHFS13}.

% However, existing treatments have either used a more liberal
% definition of linearizability and required lock-free objects
% \cite{GY11}, or used contextual refinement to characterise progress
% properties of concurrent objects \cite{LiangHFS13}. Furthermore,
% none of the treatments \cite{FORY10,GY11,LiangHFS13} consider the
% \emph{traces} of a client.

% Unlike Gotsman and Yang \cite{GY11} and Liang et al.\
% \cite{LiangHFS13}, we \emph{motivate} the required properties of
% concurrent objects by systematically reducing the proof obligations at
% hand. In particular, w

We have derived the sufficient conditions for contextual trace
refinement using the proof obligations for forwards and backward
simulation. However, neither of these conditions have been shown to be
necessary, leaving open the possibility of using weaker correctness
conditions on the underlying concurrent objects.  Studying this
relationship remains part of future work --- areas of interest include
the study of how the correctness conditions for safety of concurrent
objects under relaxed memory models \cite{DDGS15-ECOOP} can be
combined with different scheduler implementations for progress (e.g.,
extending \cite{LiangHFS13,HS11}) to ensure contextual trace
refinement. % The stepwise development of the

\subsubsection*{Acknowledgements}
We thank John Derrick and Graeme Smith for helpful discussions.
Brijesh Dongol is supported by EPSRC grant EP/N016661/1 ``Verifiably
correct high-performance concurrency libraries for multi-core
computing systems''.

\bibliographystyle{plain}

\bibliography{main,ls}

\end{document}